**Title: Causality and Scientific Inquiry: Lessons from Space Physics and Medical Sciences**

**Authors information:**

**<u>First and Corresponding author:</u>**
**Marzieh Asgari-Targhi, PhD**
Faculty of Arts & Sciences,
Department of Philosophy,
Harvard University
25 Quincey Street
Cambridge, MA 02138
Emails: marzieh.targhi@gmail.com
mtarghi@fas.harvard.edu

**Co-authors:**

**Second author:**
**Amene Asgari-Targhi, PhD**
Faculty of Arts & Sciences,
Department of Philosophy,
Harvard University
25 Quincey Street
Cambridge, MA 02138
Email: aasgaritar@gmail.com
aasgari@fas.harvard.edu

**Third author:**
**Mahboubeh Asgari-Targhi, PhD**
Harvard-Smithsonian Center for Astrophysics,
60 Garden Street
Cambridge, MA 02138
Email: masgari-targhi@cfa.harvard.edu

**Fourth author:**
**Edward J. (Ned) Hall, PhD**
Norman E. Vuilleumier Professor of Philosophy
Faculty of Arts & Sciences,
Department of Philosophy,
Harvard University
25 Quincey Street
Cambridge, MA 02138
Email: ehall@fas.harvard.edu

**Declarations of interest: none**

**Abstract**

Over the past two decades, the rapid surge in data-intensive computational techniques for statistical modeling may have had the effect of diminishing the use of applied mathematics in causal scientific inquiry. In this paper, co-authored by an astrophysicist, a mathematician, and philosophers, we assess the hazards of neglecting the branch of mathematics that constructs models to address causal questions in favor of statistical modeling alone. Causality is relevant in all branches of science and is often elucidated through applied mathematics. Here, we illuminate the idea with examples drawn from space physics and medical sciences. We examine causal questions to demonstrate how applied mathematical and statistical methods may differentiate between two fundamental facets of causality, i.e., mechanistic and difference-making. Understanding such foundational differences in causality may, in some cases, help explain discrepant or erroneous research results. Most importantly, understanding the relationship between causality and analytical approaches used in science has the potential to strengthen the rigor and reliability of scientific inquiry through optimal selection of mathematical and/or statistical methods.

**Keywords:** Astrophysics, solar corona, mathematical modeling, statistical analyses, causality, mechanistic explanation.

## 1. Introduction

We begin by outlining crucial differences between scientific knowledge and scientific data. Scientific knowledge has two distinct features amongst its other characteristics: prediction and replicability. On the other hand, scientific data are pieces of information without explanatory power or predictions. Scientific methods are the means by which the goals of natural and social sciences are achieved. Here we will focus on the significance of the distinction between scientific knowledge and scientific data in the natural sciences and how each is produced. We discuss scientific methods and their relations to foundational questions in causality. Moreover, we will illustrate that manipulating scientific data through statistical analyses does not often yield scientific knowledge and causal explanations. The above distinctions hold in other disciplines, too, such as social sciences. For example, recent research by social scientists, (Bedson et al., 2021), critiques the disease modeling methodologies in public health and social sciences. They suggest integrated modeling that better incorporates social and behavioral dynamics to help advance predictive accuracy in those fields.

With these crucial distinctions in place, we will proceed and propose a combination of multi-modeling methods in generating scientific knowledge. The following four fundamental, interrelated hypotheses are examined in this proposed paper. **First:** when reasoning causally, scientists often *intend* to understand a scientific claim about causality in a mechanistic rather than difference-making way. The distinction between these two ways of approaching causality will be further explicated shortly. Here, it suffices to mention that contemporary philosophical discourse on causation, particularly among philosophers and philosophers of science, has centered around two predominant conceptualizations of causal relationships. One approach conceptualizes causation as a physical connection between discrete events facilitated by an appropriate process or mechanism. Another approach frames causation as a form of "difference-making" among distinct events, enabling the control of certain events through the manipulation of others. For further elaboration on these theses, see (M. Asgari-Targhi, 2006; Hall, 2004; Woodward, 2003) and (Godfrey-Smith, 2010).

**Second:** unique scientific and mathematical methods correspond to the two causality types; (i) mechanistic causality is often investigated via lab-based experiments, mathematical modeling, or a combination of both, or lab-based experiments with non-inductive analysis of the logic of confirmation.[1] (ii) difference-making causality is usually investigated via statistical models as a test for counterfactual dependence or association, i.e. for A to cause B, B would not have occurred in the absence of A (see Table.1). The core functions of both types of modelling will be discussed: differential equations in mathematical modelling and probability functions in statistical modelling. We will examine their structures and how they contribute to causal discovery, particularly the types of causal relationships they can address. For instance, in mathematical modelling, causal relationships can be identified through two essential components of differential equations: time and state variables. A state variable is part of the set of variables used to represent the mathematical "state" of a dynamical system. Differential equations are fundamentally associated with the concept of time, as they model the rate of change of a quantity (commonly represented as a function) in relation to time. This intrinsic connection enables the analysis, description, and prediction of dynamic systems across various scientific and engineering disciplines.

**Third:** The choice of causal approach (mechanistic vs. difference-making) may influence the selection of research methods or variables. **Fourth:** Without adequate understanding of the differing merits of each approach to causality, scientists may more willingly rely on the evidentiary value of statistically based models instead of using other methods such as mathematical models that might provide more robust answers to scientific questions. The mechanistic inference provided by mathematical models will be shown to differ from the associational results of statistical modeling which does not always result in scientific knowledge. The difference-making approach may sometimes give scientists a false sense of confidence that A is the "handle" directly *manipulating* B, when in fact the relationship is weaker or even coincidental. This is not to suggest that statistical modeling is not important or powerful when used appropriately – clearly it is.

For instance, in the field of social sciences, there is a variety of approaches to causality and causal analysis grounded in counterfactual reasoning (Lewis, 1973a; Pearl, 2009). The counterfactual approach represents a framework that emphasizes the comparison between actual occurrences and hypothetical alternatives, this perspective primarily originates from the works of Lewis (1973a, 1973b). The core principle underlying this approach is that causal claims can be articulated through counterfactual conditionals, as mentioned in (ii) above also see (Paul & Hall, 2013).

The counterfactual approach has led to the development of several specific models (Heckman, 1974, 2000, 2005; Heckman & Moktan, 2020; Heckman & Pagés, 2004; Morgan & Winship, 2014; Neyman, 1935, 1990; Pearl, 2009; Rubin, 1974, 2006). A particularly influential variant of this approach is the potential outcomes model, which constructs hypothetical scenarios to analyze any counterfactual situation. Two salient features of potential outcomes models have significant implications for causal inference. The first aspect is the reliance on counterfactuals: An event A is considered to be a cause of effect B if and only if B depends on A in such a way that, had A not occurred (or occurred differently), B would not have occurred (or would have manifested in a different way). While this idea may seem straightforward, it is essential and, as we will explain, is not always intuitive or obvious. The second feature focuses on interventionist approach, which asserts that the only variables a scientist can consider causal are those that (a) can, in principle, be manipulated independently of other factors and (b) whose manipulation would yield observable changes in the outcome of interest. This perspective is firmly anchored in the tradition of randomized controlled trials, hereafter RCTs. Interventionism argues that randomized experiments, in which participants or units are systematically assigned to various treatment conditions, constitute the most robust approach for identifying causal effects. This viewpoint, however, is not without its critiques, notably from scholars such as (Deaton & Cartwright, 2018) and (Shelby, 2016). Consequently, any variable that, in principle, resists experimental manipulation cannot be considered causal. As we will show, this position is contentious and does not always facilitate causal discovery.

[1] An example is the case of Semmelweis's inferential procedures in establishing the cause of childbed fever (Carter, 2003; Carter & Carter, 1994, Chapters 3–4).

An alternative line of inquiry that adopts Lewis's counterfactual perspective employs directed acyclic graphs (DAGs) (Pearl, 2000, 2009). In this methodological framework, causal models are depicted as directed graphs, with vertices representing relevant variables and edges denoting the relationships of counterfactual dependence among these variables. While this approach has been effectively applied to simple illustrative examples, its applicability to complex real-world scientific problems remains underexplored. Imbens (2020) draws a clear distinction between the potential outcomes model and DAG methodologies, making the case that prevailing norms and practices in disciplines such as economics have contributed to the predominance of the potential outcomes model in that field. In section three, applications of statistical modeling for causal discovery within the scientific disciplines will be discussed. Two examples from the health sciences will be presented: one within a historical context and the other reflecting contemporary practices. Each example will illustrate the specific application of statistical modeling methodologies and their alignment with the two components of the counterfactual approach to causality: counterfactuals and interventions.

This article aims to sharpen the distinction between mechanistic and difference-making causality as a foundational concept for scientists in all fields. A better understanding of the concepts of causality should improve methodological choices by scientists and may at times explain contradictory or unsupportable research results. We emphasize that both forms of modeling, mathematical and statistical, have their proper place, as each addresses distinct types of causal questions. However, the greater accessibility and ease of application of statistical analysis can lead to its disproportionate use in practice, sometimes at the expense of deeper mechanistic understanding. The central thesis of this paper is not to privilege one mode of inquiry over another, but rather to draw attention to a prevailing methodological imbalance, specifically, the disproportionate reliance on statistical approaches across many areas of scientific research.

Scientific causal questions are frequently framed in ways that implicitly or explicitly favor statistical modeling as the most appropriate tool. As we will see in the case studies and example discussed, this tendency can yield varying outcomes. In some instances, such as the work of Hernán and Robins (2016) on hormone therapy for postmenopausal women, statistical causal modeling is applied with considerable rigor and insight, yielding invaluable results with significant clinical and public health implications. In contrast, the Keys example illustrates the limitations of relying exclusively on such methods. We argue that mathematical modeling warrants broader recognition as a legitimate and necessary complement to statistical approaches, especially in contexts where causal explanation hinges on elucidating the internal structure and dynamic processes of the system under investigation.

## 2. Scientific Methods: Mathematical Modeling vs. Statistical Modeling

By mathematical modeling, we mean approaches that use mathematical concepts and equations to represent a complex system (or a selected aspect of such a system) and the relationships between its components (A. Asgari-Targhi, 2017; Zank et al., 2012). Mathematical models can be conceptual, meaning that a few equations and parameters are used to describe the underlying dynamical properties of a system (Bruce & Pittendrigh, 1957; FitzHugh, 1961; Hodgkin & Huxley, 1952; Kronauer et al., 1982; van der Pol, 1926) or realistic models using more complex equations and larger numbers of variables and parameters (Courtemanche et al., 1998; Forger & Peskin, 2005). Realistic modeling is commonly referred to as computational modeling since solving such model equations is not feasible analytically and requires computer simulations. Mathematical models differ in structure, underlying assumptions, level of desired quantitative detail (i.e., number of parameters and variables), and quality of available experimental data for applying constraints on model variables. Mathematical modeling enables scientists to observe, analyze, and virtually manipulate a complex system *in silico* when empirical methods are not feasible (e.g., *in vivo* and *in vitro* experiments in physiological and biological systems) or when the target system is experimentally inaccessible (e.g., the solar corona or a black hole).

By statistical modeling, we mean modeling that applies statistical techniques to describe the dataset's features and study the relationship among its variables. Scientists employ such techniques to identify and discern 'causal' patterns in data and determine the subject matter for causal questions. While most researchers and scientists are familiar with the distinction between mere association, correlation, and causation, and despite the widespread use of causal concepts in the sciences, a belief lingers that causal inquiry can be reduced to statistical modeling, where the emphasis is on associations and correlations between variables.

These two types of models (mathematical and statistical) can often overlap, especially when a mathematical model's outcome requires validation with experimental data in which statistical techniques are utilized. However, one of the distinctive features exclusive to mathematical modeling, which makes the process robust in providing mechanistic explanations for causal inquiries, is the validation process. This is an intrinsic part of mathematical modeling and makes the construction process considerably longer than developing statistical models. The time it takes to build a mathematical model makes it less desirable to scholars who aim to find associations between defined variables fast.

### 2.1. Three significant differences between mathematical and statistical modeling

What distinguishes these two approaches are: (I) how they answer causal questions, what types of assumptions are employed in forming a scientific hypothesis, (II) what mathematical tools are used to establish models, and (III) if a validation and verification process is embedded within a given model or the verification part is added at the end of the analysis.

***(I) Primary assumptions.*** In designing mathematical models of dynamical systems (e.g. biological, astrophysical systems), basic implicit parameters/variables represent the scaffolding of that system. For instance, one of the early mathematical models of a biological system, the Hodgkin–Huxley model, was designed to depict how action potentials[2] in neurons are initiated and propagated. The basic features of the model represented the biophysical parts of that system, and the model studied the behavior of the components and their relationships over a certain period of time. While mathematical models do not always depict the detailed parts of dynamical systems (e.g., biological, astrophysical systems), and they have their limitations, the basic assumptions in establishing the model are crucial in identifying pathways that may connect components of that model. On the other hand, in statistical modeling, the model's design is sometimes based on generating data and the relationship between random and non-random variables within the gathered data. Thus, the primary assumptions that form such model's scaffolding either do not exist or at least not in a similar way as they do in the design of mathematical modeling.

***(II) Mathematical Tools: Functions and Equations.*** Tools that capture and describe the relationships between a set of variables within mathematical or statistical models are different and serve distinct purposes. In mathematical modelling, the non-linear differential equations and probabilistic functions are both employed. The differential equations define a relationship between the functions representing physical quantities and their derivatives representing their rates of change. However, differential equations employed in applied mathematics and those used to solve real-life problems may not necessarily be directly solvable, i.e., do not have what is known in mathematics as *closed form* solutions. Thus, solving such model equations is not feasible analytically. Instead, solutions can be approximated using different methodologies, such as *numerical methods*, which require computer simulations (i.e., computational models).

Nevertheless, the mathematical modeling approach has the potential to capture the mechanistic structure of a dynamical system, allowing for the discovery of the mechanistic aspect of causality. In statistical modeling, techniques such as curve-fitting, regression analysis, descriptive statistics, and nonparametric statistical formulation are used/applied to a dataset to study its features and the relationship among its variables, to estimate the parameters and variables of the model, and to make predictions. Statistical modeling approaches have the potential to capture the previously unknown relationship between variables in a dataset. Thus, it allows for discovering the difference-making aspect of the causal processes, but they do not always identify the mechanistic connections between the variables. One of the earliest examples of differential equations capturing causal processes over time is the Lotka-Volterra equations (Schaeffer & Cain, 2016). These equations explain the dynamics of predator-prey populations by denoting their terms to predation rate and reproductive rate. Consider the following two equations: the first one represents exponential growth, and the second one represents a balance between two effects in an equation:

$$\frac{dx}{dt} = ax - bxy \qquad (1)$$

$$\frac{dy}{dt} = cxy - dy \qquad (2)$$

[2] An action potential occurs when the cell's membrane potential (voltage difference between the inside and outside of the cell) rapidly rises and falls.

The parameters in the Lotka-Volterra system are all positive in the mathematical sense and denote the following underlying physical assumptions. The system describes the evolution of two interacting populations, a predator (for example, foxes, represented by $y$) and a prey (for example, Squirrels, represented by $x$). When there are no predators (i.e., $y = 0$), the prey population satisfies $dx/dt = ax$ the equation for exponential growth. But, when their growth rate is reduced by predation, which is assumed to happen at a rate proportional to each population. The predator equation for $y$ represents a balance between the two effects: the fox's population increases by a term proportional to the amount of food foxes (the predators) consume and decreases by a death proportional to their population. The two effects are added, and in general, each individual effect can be added.

The equations illustrate that predation directly affects the number of prey populations. The depletion in the prey population changes the size of the predator population, and reproduction restores the prey population. These differential equations capture the rate of change through time better than the probabilistic functions at the core of statistical modeling. As we will illustrate in section four through the solar corona example, that mechanism of change is discovered without narrowly focusing on intervening variables or overreliance on the counterfactuals. The Lotka-Volterra model serves as a significant example of how differential equations are essential for representing the dynamics of evolving phenomena over time. It is also interesting to know that the underlying processes are modelled probabilistically, necessitating the use of both differential and probabilistic functions in the analysis. Schaeffer & Cain (2016, p.18) point out that differential equations model plays a key role in approximating the evolution of average populations, especially when dealing with large populations**.** The rate term proportional to $xy$, they explain, can be derived from the probability that members of the two species encounter one another.

Let us unpack the relationship between mathematical modeling, particularly the use of differential equations, and mechanistic causation, especially in light of philosophical debates about the explanatory relevance of formal models. We argue that models such as the Lotka-Volterra system can and should be interpreted as mechanistic causal models when they are appropriately grounded in real-world biological processes. On the standard account of mechanistic explanation (Machamer et al., 2000), a mechanism is constituted by organized entities and activities that are responsible for producing a phenomenon. We will expand on this discussion further, but here our point is that in this framework, the Lotka-Volterra equations are not merely descriptive or predictive tools; rather, they represent a minimal but coherent mechanistic system: prey and predator populations are the entities, and the processes of reproduction, predation, and natural mortality are the activities. These components are dynamically interrelated, forming a feedback structure that captures the system's temporal and causal organization.

Each term in the system of equations corresponds to a distinct causal process. For instance, the term $ax$ reflects the intrinsic growth of the prey population in the absence of predation, while the interaction term $bxy$ captures the causal influence of predator-prey encounters leading to prey reduction. Predator growth $cxy$ and decline $dy$ are likewise causally dependent on prey availability and natural death, respectively. These are not merely correlational dependencies but mechanistic relations that encode causal productivity. As Illari (2011) and Woodward (2003) argue, mathematical models can represent causal structure when their terms are interpreted as standing for productive processes and when they support counterfactual reasoning. The Lotka-Volterra model permits precisely such reasoning: interventions on one variable (e.g., decreasing predator abundance) yield predictable effects on the other, illustrating the counterfactual dependence, one characteristic of causal explanation. But the model does more than track population-level changes over time with the difference-making apparatus, it articulates how those changes are brought about through internal system dynamics, i.e., articulations of mechanism.

Having said the above, we emphasize that mathematical formalism, in itself, is neither causal nor mechanistic, but acquires those explanatory roles when the mathematical terms are systematically mapped onto real-world processes. In the case of the Lotka-Volterra model, the differential equations do not merely describe changes in population size but embody an abstract representation of causal processes such as predation and reproduction. Thus, the model meets the central requirements of mechanistic explanation: it identifies the relevant entities and their productive interactions, organizes them into a coherent causal structure, and articulates how interventions on one part of the system generate systematic changes in others as the mechanism unfolds over time. For a thorough investigation of differential equations as the core function of causal mechanistic discovery see (M. Asgari-Targhi, 2025). At the same time, we recognize that recent philosophical work has called attention to limitations and challenges for mechanistic interpretations of mathematical and dynamical models, especially when models are highly abstract or non-decompositional.

For example, Carrillo & Knuuttila (2023) argue that many abstract models, including network models or dynamical systems, do not align easily with the mechanist's decompositional ontology, because their epistemic strategy does not depend on identifying detailed components and activities, but instead focuses on relational structure or coarse-grained behavior. In contexts like these, even if the mechanist can show that certain mechanisms are implicit, the explanatory power comes more from the dynamics over time and from how interventions affect the system than from specifying every internal component. Our use of the Lotka-Volterra model, in contrast, is not merely abstract: it explicitly maps variables to biological agents and processes; it shows how altering one variable (e.g. prey population or predator mortality) over time leads to specific changes in the other; and exhibits clear causal feedback loops. Thus, while we acknowledge the insights when mechanistic explanation becomes strained by abstraction, we maintain that in the case of Lotka-Volterra, and other models similar to Loka-Volterra, the mechanisms are sufficiently concrete, the interventions interpretable, and the temporal dynamics fully represented to satisfy a robust mechanistic causal account.

Carrillo & Knuuttila's (2023) critique does not apply to many models used in applied biology. For example, the Hodgkin–Huxley model in neurophysiology (Hodgkin & Huxley, 1952), compartmental SIR models in epidemiology (Kermack & McKendrick, 1927; Hethcote, 2000), cross-bridge models of cardiac muscle contraction (e.g., Mashima & Kabasawa, 1984; Taylor et al., 1992), a range of cardiac physiology models extending the Hodgkin–Huxley framework, and a previously developed model of cardiac electromechanics adapted and applied by Ashikaga and A. Asgari-Targhi (2018) all exemplify mechanistic approaches that aim to capture the structure and dynamics of the systems they represent.

The Ashikaga and A. Asgari-Targhi's (2018) adapted model of cardiac electrical activity employs information-theoretic tools to locate points of phase transition in heart tissue, linking changes in signal transmission to physiological wavebreaks (points where a propagating electrical wavefront in the heart fails to continue smoothly). Despite its mathematical sophistication, the model remains grounded in identifiable biological components and processes, thereby exemplifying how dynamical systems can reveal mechanistic structure through time-evolving interactions and perturbation analysis.

These models are typically formulated using systems of differential equations and explicitly encode underlying causal mechanisms. In line with philosophical accounts of mechanistic explanation (e.g., Craver, 2007; Machamer et al., 2000), they do so by mapping biological entities and their interactions onto mathematical variables and formal structures, thereby representing the organized activities responsible for producing a phenomenon. They are also used to trace causal dynamics under experimental manipulation and across time, illustrating how mechanistic modeling can capture feedback-driven, time-dependent physiological processes in ways that statistical approaches alone often cannot.

***(III) Validation process.*** The validation process is embedded in mathematical models. The model is validated against independent data sets according to its main objectives. Model validation is an integral part of the modeling process, and it involves an iterative process of *integrating modeling* and *experimental validations* (A. Asgari-Targhi & Klerman, 2018). This process usually unravels the unknown properties of the system under study, and it may lead to the discovery of new frontiers in the field employed, thus aiding the development and design of new experiments. In statistical modeling, practices such as cross-validation, likelihood analysis, and model selection are widely employed. Nonetheless, the challenge of replicability remains a significant concern (Ioannidis, 2005; Ioannidis et al., 2001; Lim et al., 2012). Unlike mathematical modeling, where validation is inherently integrated into the design process, in statistical modeling, the validation phase is often treated as a separate and distinct component.

With the above distinctions between the two modeling, we will proceed and provide concrete examples in the following section. We will depict how the causal differences between mechanistic and difference-making approaches correspond to the distinction between mathematical and statistical modeling.

## 3. An Analysis of Two Concepts of Causality and Their Correspondence with Mathematical and Statistical Modeling

Explanatory knowledge is a broad topic rooted in the fundamental concept of causation (for detailed treatment of explanations and causal explanations, please see(Achinstein, 1983; Salmon, 1984; Skow, 2016; Strevens, 2008). Philosophical work on causal concepts has revealed a well-documented ambiguity in how causal concepts are used, an ambiguity which may give rise to distinctive issues in methodological choice in scientific investigations. Sometimes, we humans think of cause-effect relationships as grounded in mechanisms or processes by which an effect may be physically traced back to its causes; but sometimes, we think of causes as “handles” by which we can potentially manipulate their effects. Neither way of thinking can be easily reduced to the other, and both are indispensable (M. Asgari-Targhi, 2006; Hall, 2004), see also Woodward (Woodward, 2003) and (Kutach, 2014).

When studying the causal structure within a specific domain, one of the preliminary tasks is to identify a set of variables—referred to as parameters, characteristics, or traits—that effectively represent various facets of the systems under investigation. The subsequent objective is mapping causal relationships among these variables. A point that is both very important and easy to overlook is that these causal relationships themselves have two quite distinct aspects.

On the one hand, a given variable **A** might be a *difference-maker* for another variable **B**: that is, it might be that a range of targeted interventions on the value of **A** would each have led to **B** taking on some corresponding value. Such difference-making relationships show up constantly both in ordinary life and in scientific contexts. For instance, whenever a scientist makes a measurement of some parameter, she relies on difference-making relationships: for the measurement technique to count as reliable, it needs to be the case, minimally, that the value of the parameter being measured makes a difference to the measuring apparatus's recording. In short, difference making relationships are ubiquitous everywhere and developing our knowledge of them is central to the scientific enterprise.

On the other hand, variable **A** might be *connected by a mechanism* to variable **B**: that is, it might be that when **A** takes on a certain value, that initiates a sequence of events, each a cause of the next, that culminates in **B**'s taking on some value. Again, causal mechanisms are ubiquitous, easy to recognize in both ordinary life and in scientific contexts. In each of the examples in the previous paragraph, the very *fact* that the value of a variable depends on the value of another is mediated by one or more connecting causal mechanisms. Here is another, utterly prosaic example: Suppose that Suzy, a young vandal, throws a rock at a window, and the window breaks. If she had not thrown, the window would not have broken; so we have a difference-making relationship. But in addition, there is a causal mechanism leading from the throw to her breaking: the rock's flight through the air. Once you start looking for them, you will find mechanisms are ubiquitous, too and developing our knowledge of them is *also* central to the scientific enterprise. For a comprehensive treatment of these issues, see (Hall, 2004; Paul & Hall, 2013).

Several observations about mechanism and difference-making are in order.

First, it is plausible that there can be no difference-making *without* mechanism: when the value of one variable B depends on the value of another variable A, there must be some reason why this dependence obtains, and that reason will trace back to some causal mechanism or mechanisms connecting A with B. There may be exceptions; for instance, quantum entanglement as manifested in the experimental violations of Bell's inequalities is an example of dependence without connecting mechanism (Maudlin, 2014).

But, second, this connection need not take the simple shape of a single causal process with A at one end and B at the other. As an example, suppose we modify the example of Suzy's rock: this time, her friend Billy rushes to intercept her rock mid-flight, so the window does *not* break. If he had not acted, the window would have broken; so there is a difference-making relationship between the window's state and his decision to try to intercede. But it is mediated not by a single causal process, but by the fact that his decision initiates a process that interferes with a second process which (had it not been interfered with) would have led to the window's breaking (Hall, 2004).

Third, while it's plausible that there can be no difference-making without mechanism, there can certainly be a mechanism without difference-making. As yet another variant on our example, suppose that Billy in fact wants the window to break, and is all set to pick up a rock to throw at it; but, seeing that Suzy has already done so, he doesn't. If Suzy had not thrown her rock, the window would have broken all the same; so the difference-making relationship present in the first version of this example is now absent. But the very same mechanism remains in place. We might compare here a familiar case from biology: a given gene might regulate some biochemical process via some mechanism, even though, had that gene been absent or non-functional, some *other* gene would have regulated the process instead. In short, any time we have a "backup" mechanism, difference-making relationships will be absent or attenuated.

Before we examine some methodological observations through the two examples from medicine in this section, it is important to highlight the extensive debates surrounding causal structures in philosophical literature. These discussions often employ causal models that incorporate structural equations and probabilistic interpretations of causal relationships, particularly framed in terms of difference-making. Substantial body of literature has been written on probability, the conceptual understanding of causality, and the nature of variables. These mathematical theories of causality center around statistical interpretations of causal relations. Salmon et al. (1971) presented an important interpretation of causal relations in terms of statistical relevance, or conditional dependence relations. Consider a class or population D, an attribute E will be statistically relevant to another attribute F, if and only if $P(F|\mathrm{D.E}) \neq P\ (F|D)$, that is, if and only if the probability of $F$ conditional on D and E is distinct from the probability of $F$ conditional on D alone.

The intuition behind the statistical relevance model is that statistically relevant properties, attributes, etc., convey explanation, and those not statistically relevant do not convey such explanatory information. The statistical relevance model assumes that explanations must convey causal relationships and that causal relationships are completely captured by conditional dependence and independence of statistical relevance relationships. These two assumptions are controversial, and the latter is clearly inaccurate as there are numerous statistically relevant phenomena with no causal relationships. Salmon (1984) moved away from describing causal relationships in terms of statistical relationships and introduced a new theory of explanation, named a causal mechanical model of explanation, which demonstrates the explanatory properties beyond statistical relevance. This model is similar to mechanistic theories of causality explained in the next section.

There are also statistically based theories of causality that make even stronger assumptions than those made by Salmon, which connect causation and probability. Spirtes et al. (1993, 2000) put forward a theoretical approach called "Mathematical Causal Models." These works, along with Pearl's formalisms (Pearl, 2009) have contributed to the debates in the philosophy of causation. They utilized a pre-existing formalism, parametrized directed acyclic graphs, known as Bayesian networks, to represent the probability constraints and causal claims of different classes of statistical models that are used to explain causality in continuous and categorical data. A causal model predicts the behavior of a system using probabilistic functions. It implies the truth value or probability of counterfactual assumptions about the system and predicts the effects of interventions. The model consists of variables and their probabilistic dependence or independence on one another.

Probability is also employed to represent the uncertainty about the value of unobserved variables in any given case or the distribution of variable values in a population. There are practical concerns with these approaches, and they have not been implemented in scientific cases. The structural equation models (SEMs) can be deterministic or probabilistic with random errors. The deterministic SEMs rest heavily on the counterfactual interpretations of the difference-making aspects of causation and are used in discovering the actual causation using counterfactual dependence. Whereas probabilistic SEMs employs probability distribution over the variables, hence on the interventionist aspect of difference-making causality.

The SEMs that rely on counterfactual dependence face serious issues and are unable to account for preemption and overdetermination cases. In a preemption case, one event negates the causal relevance of another event, even though the second event would have been a cause if the first had not occurred. In contrast, the overdetermination case arises when two events occur simultaneously, and either of them would have been sufficient on its own to cause the outcome (see Hitchcock, 2007; Paul & Hall, 2013, Chapter 3).

There are mathematical tools such as the Markov Condition,[3] formulated and developed by Pearl & Verma (1992), and Pearl's do-operator (Pearl, 2009), both of which have attracted significant philosophical interest. The Markov Condition, when taken as an axiom, underpins key results in causal graphical models, most notably, the correspondence between d-separation and conditional independence relations established by Pearl and Verma (1992). Pearl's do-operator (2009), developed within the framework of probabilistic structural equation models (SEMs), formalizes the notion of intervention within a given system and has become especially influential in discussions of causal inference. Here, we focus on Pearl's do-operator. Nevertheless, despite these advances, conceptual challenges such as preemption and overdetermination remain unresolved within these formalisms.

A conditional probability of $P(B = b \mid \mathrm{A} = a)$ is the probability that $B$ will take the value $b$ , provided that $A$ has been observed to have the value $a$. However, we are usually interested in predicting the value of $B$ that will result if we intervene to set the value of $A$ equal to some particular value $a$. Pearl (2009) defines $P(B = b \mid \mathrm{do}\ \mathrm{A} = a)$ to characterize this conditional probability. Notice the do $(\mathrm{A} = a)$ is not an event in the original probability space. Assumptions about the variable's value will give us information about its causes and other effects of those causes. But when we intervene, we change the normal causal structure, making a variable take a value it might not have taken if the system were left alone.

The antecedents of structural counterfactuals are being realized by interventions. Pearl (2009) distinguishes interventions from hypothetical statements. For Pearl, interventions are in the indicative mood and are represented by the do operator, and they are actually performed. Epistemologically, the differences between observations and interventions are important, but there are issues and problems with Pearl's characterization of interventions. For instance, it concerns the scope of the requirement that an intervention on a variable only targets that and leaves untouched all other mechanisms besides the processes that had set the value of that variable prior to an intervention. This prerequisite raises a circularity problem; that is, our intervention on the variable must keep intact the causal mechanism that connects cause to its effect, and if we intend to find out via our intervention what it is for that cause to bring about its effect, then we use the intervention *itself* to do so. Woodward (2003) has put forward a much more polished concept of intervention that does not refer to the relationship between the variable intervened on and its effects and avoids this problem.

Woodward's interventionist account of causation has exerted considerable influence, especially within philosophy of science, cognitive science, psychology, biology, and the social sciences. By grounding causal explanation in notions of manipulability and counterfactual dependence, Woodward provides a robust conceptual foundation for understanding how causal claims function both in scientific practice and in human cognition. The specifics of this topic extend beyond the boundaries of this paper. It is important to highlight that the particular philosophical frameworks developed by figures such as Pearl, Verma, and Woodward have played a foundational role in shaping contemporary theories of causation, particularly in the formal and empirical sciences. Causal discovery methods have been applied extensively across a wide range of scientific fields, from gene expression analysis to epidemiology. Formal accounts of causality developed in philosophy offer valuable conceptual tools, often operating at a high level of abstraction and illustrated through stylized or idealized scenarios that clarify foundational principles. For example, while Pearl's formal framework, particularly his development of structural causal models (SCMs) and the do-calculus, has had a transformative impact on fields such as epidemiology, computer science, and economics, its initial presentation was highly formal and often demonstrated through simplified examples.

---

[3] The Markov Condition is a foundational assumption in the formal representation of causal relationships, particularly within Bayesian networks and Structural Causal Models (SCMs). It states that each variable in a directed acyclic graph (DAG) is conditionally independent of its non-effects (non-descendants) given its direct causes (parents). While the Markov Condition is widely used in statistical causal inference, its philosophical significance lies in how it connects causal structure with patterns of statistical independence, thereby making causal claims empirically testable under certain assumptions (e.g., faithfulness). It is, however, not a theorem but an assumption or axiom, whose applicability depends on the adequacy of the underlying causal representation.

Pearl characterizes these as "toy problems," designed to illustrate fundamental concepts in causal reasoning (Pearl, 2002, p. 106). Similarly, the notion of "The Ladder of Causation" is introduced through idealized scenarios rather than fully developed empirical case studies (Pearl & Mackenzie, 2018, Chapter 1, p. 23). These examples, often involving just a few variables, are designed to clarify foundational concepts, and to illustrate distinctions between association, intervention, and counterfactual reasoning, and have broader implications for fields such as public health and the social sciences. Therefore, when Pearl refers to "toy problems," he acknowledges their didactic utility, while also recognizing their limitations in modeling the complexity of real-world scientific phenomena. While such abstractions are essential for theoretical clarity, they differ significantly from the modeling strategies used in complex empirical domains. Here, we acknowledge the importance and widespread application of statistical causal inference methods and emphasize throughout this article their computational power and growing influence across many fields.

Our aim is to elucidate how philosophical accounts of causation often serve as foundational conceptual frameworks for empirical modeling practices and play a pivotal role in the precise formulation of causal questions. The two facets of causation—mechanistic and difference-making—frequently give rise to distinct mathematical tools and formalisms. Mechanistic approaches typically motivate the use of dynamical, differential equation–based, or simulation-based models that aim to capture the underlying structure and temporal evolution of physical systems. In contrast, difference-making approaches tend to align with statistical and probabilistic methods, particularly in the form of structural equation modeling and causal discovery algorithms. As we will note later in this section, elements of philosophical and computational frameworks have been incorporated into causal inference methodologies, such as Hernán and Robins' framework for causal inference (Hernán & Robins, 2016).

While the difference-making approach has been extensively examined and emphasized within the applied sciences, the mechanistic dimension calls for a firmer grounding in the philosophical foundations of causality, foundations that, in turn, demand deeper conceptual clarification and emphasis.

In scientific practice it is much easier to obtain knowledge of difference-making relationships than to obtain knowledge, especially detailed knowledge, of connecting causal mechanisms. A well-designed series of randomized trials might establish to a high degree of confidence that a certain drug is effective at curing a certain disease, i.e., whether the drug is administered is a difference-maker for whether the disease remains present. But it can do so while leaving researchers in the dark as to *why* it is effective – i.e., what the mechanism or mechanisms by which the drug effects its cure are? But detailed knowledge of mechanisms is typically much more scientifically *valuable*. As an example, consider the progress of chemistry since the mid-19th century: while this discipline understandably and necessarily needed to *begin* by focusing on difference-making relationships (e.g., the difference to reaction products made by different choices of reactants), it reached maturity only when it began to systematically uncover and develop theories of reaction mechanisms. A similar story could be told about the early days of what became modern physics or modern genetics.

We think that two cautionary lessons emerge from these remarks, and it is these lessons that we will explore in the remainder of the paper. The first is that too narrow a focus on difference-making relationship can lead to *stagnant science*: researchers, especially in areas where large amounts of data can be gathered, rest content with using various well-known statistical techniques to produce evidence of difference-making causation, overlooking the singular importance of taking on the much more difficult but much higher payoff work needed to uncover causal mechanisms. Second, there is the danger of *sloppy science*: because it is comparatively easy to mine data so as to uncover prima facie evidence of a difference-making connection (easy, that is, compared to the work needed to establish mechanism), researchers may, in some cases, overstate their results, and overlook the need for careful follow up analysis to make sure that the prima facie evidence holds up under scrutiny. Here too a focus on uncovering causal mechanisms can help, because it provides an independent way to check that the difference-making connection seemingly revealed in one's analysis is in fact real.

If the difference-making approach is privileged, we will see causal inference as fully warranted only if the difference-making variable "handle" underlying the inferred causal association can be identified. The view that difference-making needs to be verified for a causal inference to be secured may temper with the assessment of evidence for or against the new theory. Overemphasis on statistical modeling can lead researchers to overlook the crucial need for evidence of mechanisms. The history of science is dotted with illustrative examples.

One well-known example of a causal scientific question that in fact needed mechanistic evidence but was dealt with only using difference-making statistical modeling is the Seven Countries Study (Keys, 1970). This research sought a causal link between the consumption of saturated fat and coronary heart disease. From 1958 to 1964, Keys, a nutritionist and his team collected data on the lifestyles, diet, and health of 12,763 middle-aged men comprising 16 cohorts in seven countries: Finland, Greece, Italy, Japan, Netherlands, the United States, and former Yugoslavia. Key's framework was to establish a causal link with variables such as blood cholesterol and diet and their relationship to cardiovascular risk. Keys (1970) showed a correlation between intake of saturated fats and death from heart disease and seemed to prove Keys' hypothesis. Soon the Seven Countries Study became the basis for an avalanche of public policies and guidelines in the US and around the world, alerting to the dangers of consuming saturated fats e.g., (Kromhout D, 1999; Kromhout D, Menotti A, 1994).

However, despite its impact and subsequent supporting papers by its original authors, it ultimately proved to be a classic example of difference-making approach with testing a hypothesis based on questionable assumptions surrounding the causal connection between diet and heart disease. Keys and his team are now often criticized for not providing scientifically valid reasons for the countries chosen for study.[4] Keys' fundamental oversight was that the causal question he was trying to answer required a mechanistic approach (why does saturated fat cause coronary heart disease and how does it?), not difference-making (is there a correlation between consumption of saturated fat and coronary heart disease?). He had a large number of variables with a vast dataset; correlations of any sort and at any level would be possible. The mechanistic approach would have required him to explain how saturated fat is metabolized in the human body and what other factors contribute to coronary heart disease.

The primary limitation of the study was methodological, particularly in the formulation of its hypothesis. The dietary recommendations were derived from epidemiological studies, which gather data on health and behavior to identify patterns. Historically developed to study infections, Keys and his successors adapted it to the study of chronic diseases (Keys, 1957). However, unlike most infectious diseases, chronic diseases typically require decades to manifest and are influenced by numerous dietary and lifestyle factors. The complexity of isolating these variables has been frequently highlighted by critics of the seven countries study e.g., (Bassler, 1994; Mann, 1993). To reliably identify causes, a higher standard of evidence is required, and many health scientists now employ randomized RCTs to achieve a better standard. Randomizing the study population into cohorts has the potential to control for innumerable known or unknown variables and isolate a single study variable for its effect on an outcome. RCTs are one of the ways to conclude with some confidence that A is responsible for B in the absence of identifying a clear mechanistic pathway connecting cause A to effect B. For a philosophical critique of RCTs see (Berwick, 2005; Deaton & Cartwright, 2018; Greenhalgh et al., 2015; Ioannidis, 2016).

Although Keys had shown a correlation between coronary heart disease and saturated fat, later Menotti et al. (1999) re-examined the data, and found that the food that correlated most closely with deaths from heart disease was not saturated fat, but sugar (Kromhout et al., 1989; Menotti et al., 1999). The primary supporter of the idea that sugar was the dietary factor promoting heart disease was Yudkin (see Yudkin & Morland, 1967). Yudkin warned of the dangers of sugar consumption, how it is metabolized, and its effects on the human body (Yudkin, 1972b, 1972a, 1974). Yudkin focused on the mechanisms of sugar digestion in the human body and the harm it causes to the human body, particularly the liver and heart. Once the mechanism was discovered and understood via his lab experiments, he illuminated the greater toxicity of excessive sugar consumption compared to fat.

Keys' research, by no means fraudulent or ill-intentioned, was mistaken in interpreting the connection between the two correlated variables, coronary heart disease and the consumption of saturated fats (Keys, 1970) as causal. The disagreement, controversy, and aftermath, i.e., the dominance of Keys hypothesis for three decades, are part of the social dimensions of science and how they affect scientific impact and reputation outside the scope of this paper. We can confidently draw a parallel between the methodologies the two scientists used and the two concepts of causation here. Keys’ method of causal discovery was the difference-making investigated through statistical modelling, Yudkin’s was mechanistic causal explanation, and he employed lab experiments.

---

[4] For instance, some have argued that Keys excluded France and what was then West Germany from his study because those countries had low rates of coronary heart disease despite high saturated fat diets (de Lorgeril, 2002; Ravnskov & Ascherio, 1994; Thom et al., 1985; Williams, 1995). (Keys, 1980) defended his approach by stating that his cohort selection, though non-random, was based on practicality and dietary variation. (Keys, 1970, 1980; Kromhout and Menotti, 1994).

Here, we aim to look deeper at the impact of causality and focus on the conceptual conflations surrounding causal scientific questions that can give rise to a dominance of the use of one method over the other. In many areas of science, in particular, public health and medicine, authors and editors often refrain from explicitly acknowledging the causal goal of research projects; instead, they refer to associational estimates, even though causal inference is at the heart of their scientific inquiry. For example, Hernán (2018) identified a widespread reluctance in the public health literature to use the word "causation" or, as he puts it, the "c-word". Hernán and Robins (see Hernán et al., 2008; Hernán & Robins, 2016, 2020) made significant contributions to public health and medicine by addressing existing gaps and formalizing the concept of causality through statistical methodologies grounded in the potential outcomes framework. Their work extensively utilized large observational databases[5], commonly referred to as big data. In comparison to the approach employed by Keys, Hernán and Robins' application of causal inference on big data demonstrates a more substantial impact on research outcomes.

Hernán & Robins (2016) employed statistical methods to analyze large observational datasets to evaluate the effects of hormone therapy (HT) on postmenopausal women. Their research aimed to "emulate" a randomized controlled trial by meticulously accounting for potential confounding variables. Generally, inquiries concerning comparative effectiveness or safety should be addressed through rigorously designed randomized experiments. However, when conducting randomized trials is not feasible due to ethical considerations, time constraints, or other logistical barriers, public health scientists often resort to the analysis of observational data. Causal inference derived from large-scale observational databases can be perceived as an effort to mimic a randomized experiment, termed the target experiment or target trial, which would ultimately respond to the research question of interest. When the aim is to guide decision-making among multiple intervention strategies, it becomes essential to evaluate causal analyses of observational data in relation to their efficacy in emulating a specific target trial.

In their research paper, Hernan & Robins proposed a framework for comparative effectiveness research leveraging big data, which explicitly delineates the target trial. This framework integrates counterfactual theory to facilitate the comparison of sustained treatment strategies, organizes analytical approaches, provides a systematic process for critiquing observational studies, and aids in mitigating common methodological challenges. Their findings indicate that while HT can alleviate menopausal symptoms, it is also associated with an increased risk of certain health complications, such as breast cancer, mainly when therapy is initiated later in life or administered over prolonged durations. These results underscore the critical importance of assessing individual risk factors before the commencement of hormone therapy. Hernán & Robins (2016) employed causal inference methodologies and utilized extensive observational datasets to assess the impact of estrogen plus progestin hormone therapy on the five-year risk of breast cancer in postmenopausal women.

[5] The term "big data" is more effective than "large observational databases" and offers advantages over small data due to its vast size and accessibility, which help researchers mimic target trials more easily. However, Hernan & Robins (Hernán & Robins, 2016) emphasize the importance of recognizing the limitations of large observational databases. Epidemiologists are concerned when big data is portrayed in the media as a replacement for randomized trials. To classify big data as research-grade, processes for harmonization and standardization must be implemented, considering factors like clinical workflows, coding practices, and software updates. This requires a deep understanding of the dataset and may involve costly validation studies, alongside indirect validation methods such as consistency checks and comparisons across different data sets.

Their study offers a more robust and scientifically rigorous instance of applied statistical modeling in research for two primary reasons: a) the difference-making approach to causality is particularly adept at addressing specific research questions, especially those that investigate the effects of targeted interventions and provide guidance for decision-making among various intervention strategies; and b) the formalism of causal analysis is thorough, avoiding the inference of causal connections solely through correlated variables and it does not attempt to address unrealistic mechanistic causal relations.

## 4. Two Case Studies from Mathematical Biology and Astrophysics: An Analytical Perspective

This section examines how scientists validate causal hypotheses using mathematical tools to identify mechanisms and how the scientific community accepts these hypotheses. Darden (1986, 1991) argues that searching for mechanisms is essential to scientific discovery. Historical evidence shows that scientists use strategies to uncover causal pathways. Machamer et al. (2000) assert that both the philosophy of science and biology should focus on how scientists structure their research around mechanistic explanations. Philosophers Bechtel & Abrahamsen (2005) define a mechanism as a structure that performs a function through its component parts, operations, and organization. They explain that the coordinated functioning of a mechanism is responsible for specific phenomena, as observed through changes in the properties of its parts over time (see Bechtel & Abrahamsen, 2005; Machamer et al., 2000).

Mechanistic explanations describe the organized collection of parts and operations that account for physiological regularities. Key characteristics include the phenomena, the involved parts, their relationships or potential causes, and the observed patterns of change. Mechanistic explanations in biological processes are complex and multi-faceted. Scientists observing complex systems cannot always identify which features contribute to specific behavioral patterns. Bechtel & Abrahamsen (2010) noted that discovering mechanisms is aided by decomposition and localization. Their flowchart for understanding scientific mechanisms begins with defining the phenomenon and its characteristics. Decomposition can be structural, which breaks the system into simpler components, or functional, which rephrases the phenomenon into a series of simpler behaviors or operations.

To identify a mechanism through functional decomposition, scientists first localize individual operations in corresponding parts, then simultaneously show that the operations are realized in the system and that the details with which those operations are identified are working parts of a mechanism. Bechtel and Richardson (2010) added that scientists deploy excitatory and inhibitory experiments to obtain detailed information about the mechanistic process. Bechtel (2011) further coined the term “dynamic mechanistic explanations” that encapsulate the two central strategies in mechanistic explanation: the downward-looking decomposition and holistic act of recomposition and the mathematical modeling’s role in providing the latter process. These explanatory strategies begin with decomposing the mechanism into parts and operations using a range of laboratory-based experiments.

The mechanism is reconstructed using mathematical models where variables in differential equations represent the properties of its components. This recomposition involves detailing the spatial arrangement of components and the temporal organization of functions, aiming to produce the phenomenon that requires explanation. Bechtel & Abrahamsen (2013) describe mechanistic explanation in computational biology, providing case studies on circadian rhythms. Biologists first identify essential components of the molecular mechanisms involved and then use mathematical modeling to determine if these mechanisms can produce sustained oscillations. Without mathematical modeling, scientists cannot specify how oscillations in individual neurons synchronize within networks. Therefore, such models were used first to investigate the dynamics of mechanisms in the control of circadian rhythms and then evaluate the process in which different possible network architectures produced the observed synchronized activity. The above-mentioned mechanistic approaches in the philosophy of science focus on molecular biology, basic cell biology, and neuroscience. Mathematical modeling plays a crucial role in these fields through the latter addition of the mechanistic explanation (ibid), that is, the recomposing process. As we will see in the next section, mathematicians use differential equations and their derivatives to identify causal pathways from one variable to another. Some in the philosophy of science, for instance, Craver (2006) and Matthiessen (2015) dispute whether mathematical modeling can generate mechanistic explanations. Matthiessen claims that the deployment of mathematics contradicts the visual and pictorial methodological norms in biology, and the explanations produced by mathematics are not mechanistic (Matthiessen, 2015, p. 3).

In light of the above critiques (e.g., Craver, 2006; Matthiessen, 2015), it is important to clarify how we understand the role of mathematics in generating causal explanations. Rather than treating "mathematics" as a monolithic tool, we differentiate between two distinct forms of mathematical tools, mathematical modeling and statistical modeling, each aligned with a different philosophical conception of causality.

Mathematical models, particularly systems of differential equations, are frequently employed within a mechanistic framework to represent the productive continuity of causal processes over time. These models capture the dynamic rates, dependencies, and structural organization that characterize a system's internal mechanisms. Unlike statistical models, which identify patterns of association or conditional independence, mechanistic models can explain how components interact causally to produce observed phenomena. Crucially, such models are not merely idealized abstractions; they are grounded in empirically identifiable entities, properties, and relations. For example, the mentioned Lotka–Volterra equations model predator–prey population dynamics by representing biologically meaningful parameters like reproduction and predation rates. Similarly, the Hodgkin–Huxley model of neuronal action potentials (Hodgkin & Huxley, 1952) uses equations tied directly to biophysical quantities such as membrane potential and ion channel conductance. In both cases, mathematical formalism functions not simply as a predictive tool but as a representational medium that encodes causal interactions over time within structured systems.

Therefore, by distinguishing between these two modeling traditions, mechanistic and statistical, and linking each to a corresponding facet of causality (productive continuity vs. difference-making), we argue that mechanistic explanations are fully compatible with certain forms of mathematical modeling, particularly dynamical models. These models are integral to constructing and expressing mechanistic explanations, especially in domains such as physiology, neuroscience, and systems biology. This clarification also addresses concerns raised by Craver and Matthiessen by showing that the use of mathematics in mechanistic explanation need not contradict the visual or diagrammatic norms of biological reasoning. Instead, it can complement them by offering a formal, quantitative account of the same causal mechanisms that those norms aim to depict.

As noted in Section 3, we included an illustrative example demonstrating how statistical analysis can be appropriately applied within a difference-making framework, namely, Hernán and Robins' (2016) study on hormone therapy in postmenopausal women (pp. 13–14). Thus, our aim is not to promote one modeling paradigm over another, but to interrogate a persistent conceptual conflation: the assumption that most causal questions in science are best addressed through the difference-making framework, largely due to the computational strength of statistical modeling. Instead, we call for methodological clarity and balance, recognizing the epistemic contributions of both approaches when causal questions are properly framed. To deepen this contrast and demonstrate how it plays out in scientific practice, we now turn to a case study from biology, where both modeling approaches, mathematical and statistical, have been applied to uncover distinct aspects of causal structure.

**4.1. Example/Case study 1: Biology.** In the study of biological systems, the distinction between mathematical and data-driven statistical modeling corresponds to the mechanistic depiction of the causal relationship and the difference-making depiction, respectively (see Table 1). Highlighting this distinction matters for translational or clinical applications. Mathematical modeling that is based on a biological system and how it is linked to experimental data can provide insights into the underlying physiology. This approach fits well with the mechanistic approach to causality, where identifying causal processes and pathways are the aims of the scientific inquiry. Furthermore, non-physiology-based mathematical models (i.e., conceptual models) can retain the system's essential features and help identify the causal relationship between the components of a model.

Models of biological systems can be divided into two main categories: dynamical models and statistical data-driven models. Dynamical models employ differential or difference equations to describe how a system evolves dynamically in time and space. Dynamical models can be deterministic, consisting of ordinary or partial differential equations, or stochastic (i.e., including stochasticity and noise that is present in experimental observations) and described by stochastic differential equations or the probabilistic chemical master equations. As described in the introduction section, dynamical models can be conceptual (FitzHugh, 1955, 1961; McKean, 1970; van der Pol, 1926) or realistic/physiology-based models (H. F. Brown et al., 1984; Courtemanche et al., 1998; Fulcher et al., 2014a; Ten Tusscher et al., 2003), and they correspond to the mechanistic approach to causality.

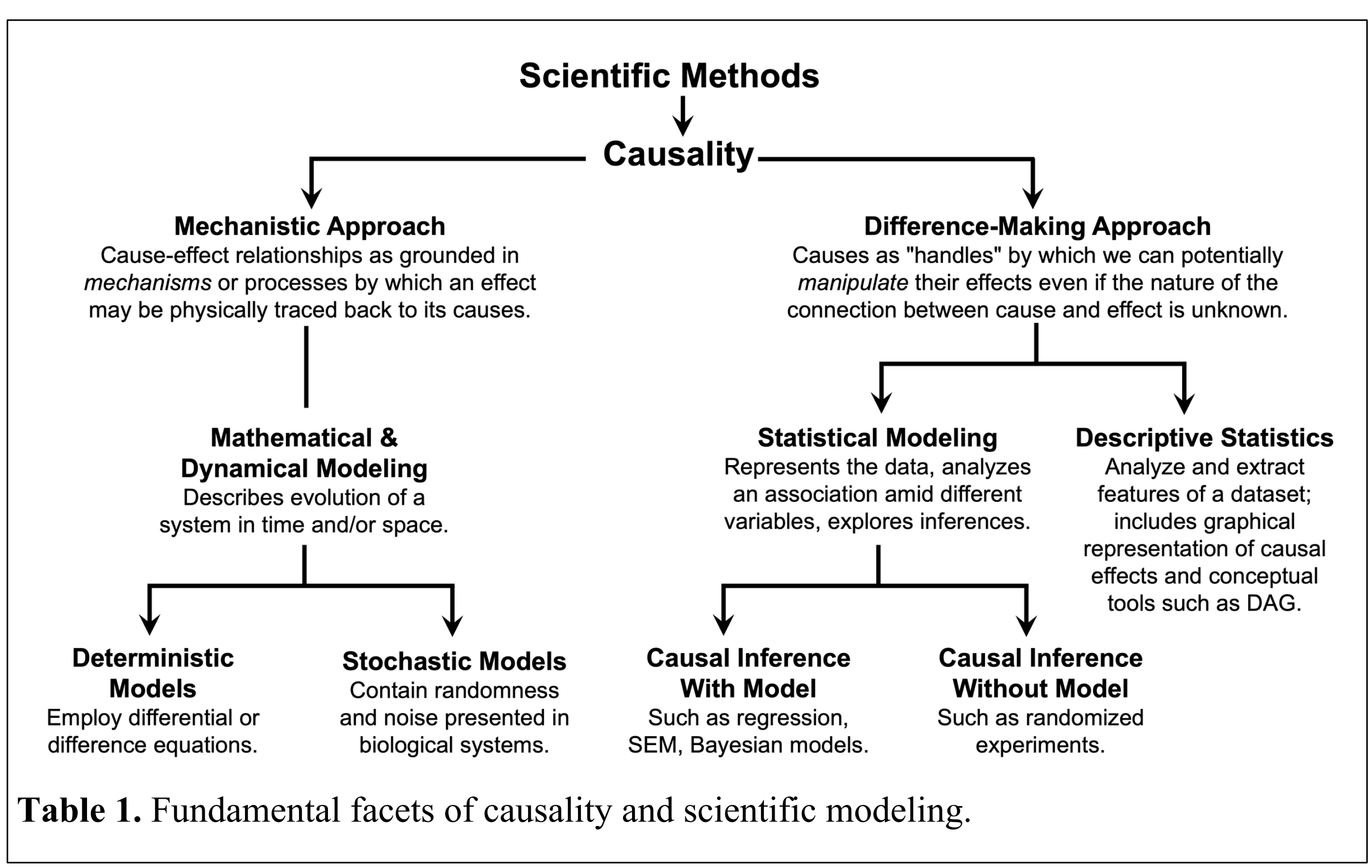

**Table 1.** Fundamental facets of causality and scientific modeling.

These models provide a valuable quantitative and mechanistic approach to understanding the inherently complex structure of biological systems, including nonlinearities of the components of the system and their relationship/interactions at specific scales, i.e., organism, multi-cellular, cellular, molecular, or genetic. Mathematical models have been successfully applied to explore many fields within physiology, including circadian physiology (A. Asgari-Targhi & Klerman, 2018; Bechtel & Abrahamsen, 2010; Kronauer et al., 1982; St Hilaire et al., 2007), sleep/wake cycle (Booth & Diniz Behn, 2014; Fulcher et al., 2014b; Phillips & Robinson, 2007), tumor dynamics, cancer research (Altrock et al., 2015; Beerenwinkel et al., 2015; Quaranta et al., 2005; Werner et al., 2013; Yin et al., 2019) and cardiac electrophysiology (H. F. Brown et al., 1984; Courtemanche et al., 1998; DiFrancesco & Noble, 1985; Luo & Rudy, 1991). In cardiac electrophysiology, models are regularly used to investigate mechanisms and treatments for cardiac arrhythmias (Callaghan et al., 2019; Pullinger & Gong, 2019). These models have identified the roles of specific ionic mechanisms, genes, and proteins in both healthy cardiovascular function and disease states (Muszkiewicz et al., 2018; Ten Tusscher et al., 2003).

Although such detailed models are ground-breaking tools for computational modeling, the level of their complexity makes their mathematical analysis rather challenging. Therefore, mathematicians often employ conceptual models that are considerably simpler than realistic models but incorporate the primary elements of complex physiology-based models to explore the system's behavior. For instance, in cardiac electrophysiology modeling, A. Asgari-Targhi (2017; Ashikaga & A. Asgari-Targhi, 2018; A. Asgari-Targhi et al., 2025) used a conceptual model of a human cardiac cell to investigate one of the potential mechanisms underlying a common irregular heart rhythm called *atrial fibrillation*, hence contributing to the understanding of the dynamical behavior of heart cells.

In situations where existing data and methods limit researchers' ability to empirically test systemic perturbation caused by unknowns and observe, analyze, and manipulate a physiologic system, mathematical modeling effectively complements experimental and clinical approaches. Mathematical models have been used to (i) investigate mechanisms and treatments for disease progression (Muszkiewicz et al., 2018; Ten Tusscher et al., 2003) (ii) explore potential ways of detecting pathological states in the early stages of a disease (Kyrtsos & Baras, 2015) (iii) investigate the potential effectiveness of new or currently available drugs that may affect disease progression (Manga & Jha, 2017; Siepmann & Peppas, 2012) (iv) develop testable hypotheses and design experiments for future therapies, (v) provide tools to reconcile contradictory results (Forger & Peskin, 2004; Gallego et al., 2006; Klerman et al., 1996), and (vi) propose new experiments based on model predictions or resolve debates.

For example, in the field of sleep and circadian rhythms, Phillips et al. (2010) developed a model of sleep physiology and circadian rhythms to investigate physiological causes of chronotype differences. The chronotype of an individual refers to that individual's tendency and preference for a sleep-wake schedule in a given 24-hour day (Adan et al., 2012). Early chronotypes tend to perform at their best in the morning across all cognitive and physical measures, while late chronotypes perform better in the evening (Facer-Childs et al., 2018; Randler et al., 2017). Chronotype influences many aspects of health, and its disturbance has been linked to various physiological and psychological disorders (Chellappa et al., 2019; Gao et al., 2019; Kivelä et al., 2018; Türkoğlu & Selvi, 2020). Phillips et al. (2010) combined models of the sleep-wake switch (Phillips & Robinson, 2007) and the circadian pacemaker (endogenous 24-hour cycles that regulate physiological and psychological processes; Forger et al., 1999), and examined the interactions between these systems. Their results suggested that chronotype is influenced by both circadian rhythm and noncircadian factors. Their modeling was an example of a mechanistic approach that helped resolve a debate over the causes of chronotype differences.

Statistical data-driven models employ statistical techniques to represent the data, analyze an association amid different variables, and explore inferences. These techniques include regression analysis to fit data, descriptive statistics to define properties of the dataset, and/or inferential statistics (Brown & Luithardt, 1999; Klerman et al., 2003; Thall, 2020). In this approach, statistical modeling techniques have been used to characterize and analyze large and complex data sets (Janes & Yaffe, 2006; Li, 2010) and identify correlations and associations between variables in a dataset. These correlations and associations can be helpful in identifying patterns in the data, generating hypotheses, and designing new experiments. However, they do not explain the causal mechanism connecting correlated variables, and thus by measuring the effects of a cause, their emphasis is on a difference-making causal approach as opposed to a mechanistic approach that focuses on identifying the causal pathways (Thall, 2020; Yang & Saucerman, 2011).

Statistical modeling approaches are commonly used to reduce the dimensionality of complex data sets to clarify and interpret these data sets, identify critical changes in data and make predictions based on the patterns present in the data (Arzani & Dawson, 2020; Brown & Luithardt, 1999; Janes & Yaffe, 2006; Klerman et al., 2003; Li, 2010; Thall, 2020; Yang & Saucerman, 2011).

**4.2. Example/Case study 2: Astrophysics** In the following case study, we will demonstrate how mechanistic and difference-making approaches perform in explaining different aspects of energy transfer into the solar corona. Highlighting such foundational differences clarifies how scientific and research questions can shape the methodology. The solar corona with a temperature of at least one million degrees Kelvin (K) is 200 times hotter than the chromosphere, its lower boundary. Within the corona, the core of active regions contains magnetically confined plasmas with temperatures above 4 million K. This apparent paradox of hotter temperatures in the higher layers is the focus of intense study. The source of energy that heats the corona and the dynamic processes that transfer energy to the corona are not precisely known.

There are two competing coronal heating theories: magnetic reconnection and wave heating (Mandrini et al., 2000). In reconnection models, the convective flows at the photosphere cause braiding and twisting of the magnetic field lines and the energy is released into the corona through a series of impulsive heating events (Parker, 1972, 1983, 1994; Priest et al., 2002). In wave heating models, the interactions between magnetic flux elements and convective flows in the photosphere produce magnetohydrodynamic (MHD) waves that travel upward along the magnetic field lines, releasing their energy in the corona (Moriyasu et al., 2004; Zank et al., 2012).

Wave models often derive from a mechanistic approach to causation analysis where a series of multiple heating events produce the energy that heats the solar corona. Magnetic reconnection has been modeled using both mechanistic and difference-making approaches (Berger & M. Asgari-Targhi, 2009).

Using this example, we illustrate how applying *only* a difference-making approach to the magnetic reconnection model overlooks other heating processes such as plasma wave heating and might favor an inaccurate conclusion that magnetic reconnection resulting from magnetic flux emergence and magnetic flux cancellation is the only process heating the solar corona. Magnetic flux emergence is, in fact, a complex process leading to the emergence of magnetic flux tubes from the interior of the Sun into the solar atmosphere (Pariat et al., 2004; Zwaan, 1985). The converse process to magnetic flux emergence is magnetic flux cancellation, where two magnetic elements of opposite polarity come into close contact and partially or totally negate each other (Harvey et al., 1999). Both magnetic flux emergence and flux cancellation release energy. The plasma temperature is often assessed using spectroscopic observations and is related to a physical quantity known as intensity $I_{hot}(x, y)$.

Figure 1(**A**) shows the spectroscopic emission intensity in the wavelength 94 Angstrom for a magnetically active area on the Sun observed on 2013 April 23. The 94 Angstrom wavelength is emitted by Fe XVIII at temperatures of approximately 6 million degrees Kelvin. The brightest loop in this region was selected for intensity measurement at a position depicted by the small red box in Figure 1(**A**). Figure 1(**B**) shows the corresponding Line-of-Sight magnetogram. The magnetic structure of this region is complex. Magnetic flux describes the total magnetic field that passes through a given area and was measured to be $20.3 \times 10^{21}$ Maxwell in this region.

**Analysis.** A study based on such observations (M. Asgari-Targhi et al., 2019) used visual inspection of the intensity images and magnetograms to categorize the magnetic complexity of 48 active regions based on the existence of sunspots, nonpotential structures (regions with electric currents), and complex magnetic structures referred to as magnetic flux emergence and magnetic flux cancellation.

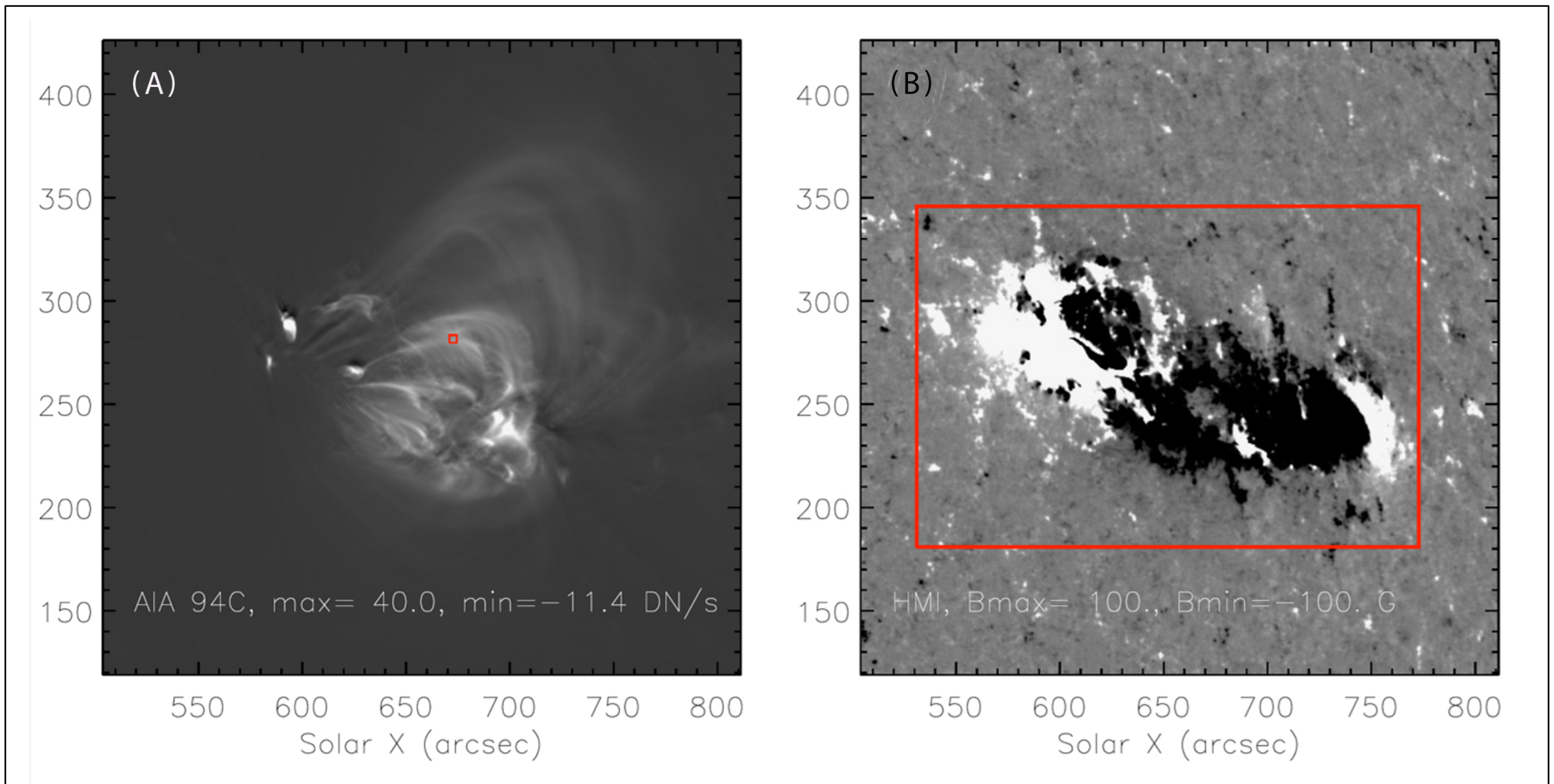

**Figure 1.** Panel (A) shows the Fe XVIII 94 Å intensity map of a region in the solar corona. This spectroscopic information is an indicator of temperature. The small red box is at the position where the highest intensity was measured. The red box in Panel (B) shows magnetogram data from the same region (Helioseismic and Magnetic Imager onboard Solar Dynamic Observatory).

Figure 2 shows the results of this study correlating spectroscopic intensity with magnetic flux, and further characterized by flux emergence and flux cancellation. Figure 2(**A**) shows the scatter plot for the intensity $I_{hot}$ of the brightest loop in active regions versus the magnetic flux. Magnetic flux measurements associated with flux emergence are depicted in blue (EC = 1); measurements associated with flux cancellation are in red (EC = 2); and measurements with neither are in green (EC = 0). The figure shows that regions with high intensities, and therefore higher temperatures, tend to have larger magnetic flux; moreover, higher intensities also tend to have flux emergence or cancellation. Figure 2(**B**) presents the Kolmogorov-Smirnov (K-S) cumulative probability distributions for observations without flux emergence or cancellation (green curve) in comparison to observations with flux emergence or cancellation (blue curve).

The K-S indicated a statistically significant difference between the distribution functions of the two samples and implied dependence of spectroscopic intensity on flux emergence and cancellation parameters.

If we focus solely on difference-making, then it would be tempting to use the correlation between temperature and magnetic complexity of these regions to conclude that magnetic reconnection is the primary driver of solar coronal heating. However, this conclusion would be simplistic and incomplete. The statistical method utilized a narrow set of parameters that favored correlation between high temperature and complex magnetic structures. It turns out that the existence of regions *without* complex magnetic structures yet *high* temperatures of 1 - 3 MK and the existence of regions *with* complex magnetic structure, yet *lower* temperatures were ignored. Taking the plots of Figure 2 at face value might lead to overconfidence in complex magnetic field configuration as the key driver of the high temperatures, despite a limited degree of correlation between the complex magnetic field and high temperature as discussed by M. Asgari-Targhi et al. (2019).

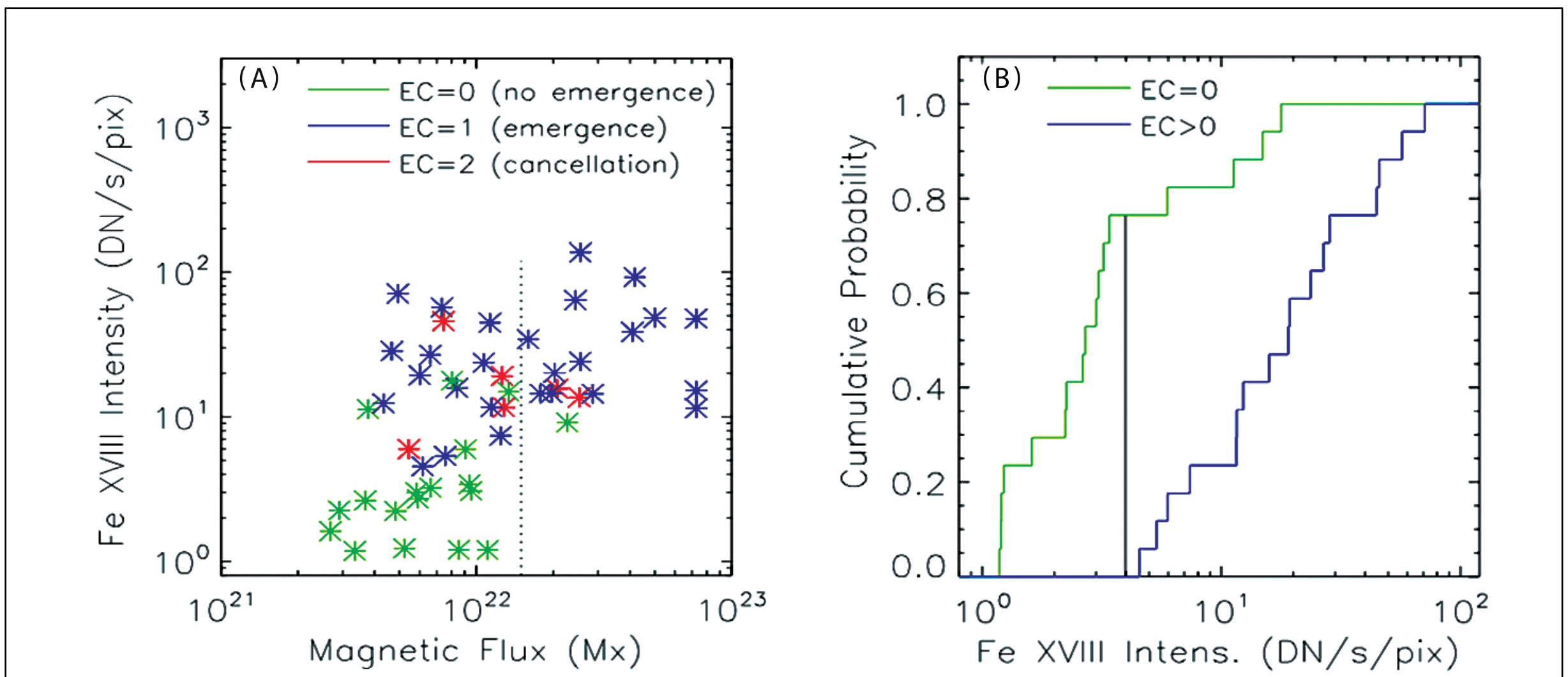

**Figure 2.** (A) Fe XVIII spectroscopic intensity $I_{hot}$ of the brightest loops as a function of total magnetic flux for the active regions is color-coded according to flux emergence or cancellation parameters (EC) (M. Asgari-Targhi et al., 2019). **(B)** The Kolmogorov-Smirnov plot compares the probability distributions of observations with no flux emergence (EC=0) versus observations with flux emergence or cancellation (EC>0) along the range of spectroscopic intensities. The wide separation of the curves was considered statistically significant.

In contrast to a purely statistical study of plasma temperature that considers a limited number of variables, mathematical modeling of the solar corona can effectively incorporate additional variables and provide a potentially more complete assessment of coronal plasma heating. In a series of papers, Asgari-Targhi and colleagues (M. Asgari-Targhi & van Ballegooijen, 2012; M. Asgari-Targhi et al., 2014; van Ballegooijen et al., 2017) have used a three-dimensional Reduced magnetohydrodynamic (RMHD) model for the propagation and dissipation of turbulence in a coronal loop. The mathematical modeling approach allowed them to take into account several parameters such as temperature, density, magnetic field strength, velocity, energy, heating rate, and the spatial and temporal variations of these parameters. Two sample equations from this work that describe the rate of change in the velocity and the magnetic field of the plasma over time are shown by equations (1) and (2):

$$\rho \frac{d\mathbf{v}}{d\mathrm{t}} = -\nabla p + \rho \mathrm{g} + \frac{1}{4\pi} (\nabla \times \mathbf{B}) \times \mathbf{B} + \mathbf{D}_v, \qquad (1)$$

$$\frac{\partial \mathbf{B}}{\partial \mathrm{t}} = \nabla \times (\mathbf{v} \times \mathbf{B}) + \mathbf{D}_m. \qquad (2)$$

where $\rho(\mathrm{r,t})$ is the plasma density, $\mathrm{p(r,t)}$ is the pressure, $\mathbf{v}(\mathrm{r,t})$ is the velocity, $\mathbf{B}(\mathrm{r,t})$ is the magnetic field, g is the acceleration of gravity, and $\mathbf{D}_v$ and $\mathbf{D}_m$ are viscous and resistive dissipation terms. The objectives of these papers were to study the dynamics of waves in the corona and to determine how interactions between plasma waves and the resulting plasma turbulence, along with changes in magnetic flux, all impact solar coronal heating.

The mathematical models described in these papers facilitated a more complete causal understanding of solar coronal heating that included regions with and without complex magnetic structures, plasma wave interactions, and wave turbulence. These mathematical models have also produced observables (M. Asgari-Targhi & Van Ballegooijen, 2012) where some of the model outputs have been validated with current solar observations(M. Asgari-Targhi et al., 2014).

Furthermore, the models have been used to predict potentially detectable variations in other parameters (van Ballegooijen et al., 2017); these predictions are subject to validation by future NASA instruments with sufficiently high spatial resolution and temporal resolutions. This more robust elucidation of solar coronal heating mechanisms would likely not be achieved by a purely statistical approach.

This case study illustrates a central philosophical point: that different causal questions require different modeling frameworks and conflating them can lead to both conceptual and methodological missteps. The statistical analysis of solar coronal heating, while methodologically sound and grounded in observational data, was limited by the causal lens it employed. By framing the problem primarily in terms of difference-making, the analysis emphasized associations between magnetic complexity and temperature while overlooking other possible causal pathways, such as wave-based energy transfer.

In contrast, the mechanistic modeling approach employed in the RMHD simulations incorporated a broader set of variables and theoretical assumptions, allowing for a more comprehensive causal narrative that integrated dynamic processes like turbulence and wave propagation. This contrast is to demonstrate that statistical and mechanistic models embody distinct causal commitments, and it is not meant to establish a general hierarchy between modeling types. Recognizing this distinction is essential, as it clarifies how the framing of scientific questions influences methodological choices and the kinds of explanations that result.

## 5. Discussion and Conclusion

This paper offers a cautionary reminder of the scientific community's attraction to big data and the use of statistical modeling techniques. We began by making the distinction between scientific knowledge and scientific data. We illustrated that statistical analyses often produce the latter rather than knowledge with predictive powers. While the numbers, results, and extracted information from statistical analyses can be *invaluable* and essential, applied mathematical models with their added qualitative properties are sometimes crucial both to deduce meaningful and insightful knowledge about a phenomenon and to correct overly hasty causal inferences.

We aimed to direct attention to foundational assumptions about causation which often lie hidden in scientific discourse. Understanding such foundational differences is vital to choosing appropriate study methodologies and may guide translational or clinical research design. Reviewing research methodologies in the context of causality concepts may account for unexpected or questionable research results. Moreover, understanding foundational differences about causation is important in assessing the strengths of statistical modeling and what it produces vis-à-vis its limitations.

With examples of applied mathematics in medicine and astrophysics, we illustrated a potentially problematic privileging of the difference-making concept of causation over the mechanistic concept in some scientific investigations. Due to the availability of sophisticated computational and statistical modeling methods, some scientists rely heavily on the technological advancement of statistical models (difference-making causality) and pay less heed to the advantages of other mathematical models (mechanistic causality), to expedite the process of publication. This has consequences for scientific inquiry and the role of applied mathematics and modeling that differs fundamentally from statistical modeling, in particular researchers can overlook the significant value of employing mechanistic models.

We demonstrated that mechanistic approach to causality that uses mathematical modeling framework has a vital role to play in applied sciences, and that, ideally, every causal claim should be backed by an understanding of mechanisms. The difference-making approach that employs statistical modeling methods cannot capture the breadth of causal questions in the applied sciences, e.g., medicine and astrophysics. The context sensitivity of tools in mathematics reflects a duality paradigm that exists in the philosophy of causality and underscores a critical role of applied mathematics in scientific inquiry.

To conclude, we do not assert the superiority of one causal framework over another. On the contrary, we have consistently distinguished between two conceptually distinct facets of causation, mechanistic and difference-making, and argued that each is best approached through a different mathematical methodology: mechanistic causation through mathematical modeling, and difference-making through statistical modeling and analysis. These frameworks are not presented hierarchically but rather as distinct and complementary tools, each suited to different epistemic and practical aims appropriate to the causal questions under investigation. They serve divergent but equally important roles in causal inquiry: mathematical models help elucidate the internal structure and dynamics of systems, while statistical models identify patterns of dependence and infer potential causal relationships.

Our central claim is not that one method should replace the other, but that the current methodological landscape places disproportionate emphasis on statistical approaches. We propose a more balanced recognition of mathematical modeling as a legitimate and necessary counterpart, particularly in contexts where capturing the structure of complex systems is essential to meaningful causal understanding.

**Conflict of interests:** Authors declare that they have no competing interests.

**Data and materials availability:** Data from case study 2: Astrophysics reported in this paper are divided into observational and mathematical models. The observational data were previously published, as were the modeling results. Analyses were conducted with Fortran, IDL, and MATLAB, and code from all studies is available upon request from authors.

**Methods:** The research case study reported in this paper in applied mathematics and astrophysics did not involve human subjects.

**Acknowledgement:**

The authors wish to express their sincere gratitude and thanks to Paul B. Shyn, MD for his invaluable feedback on earlier drafts of this manuscript. We are also thankful to Gary Zank, Professor of astrophysics for his constructive and insightful comments and suggestions.

**Authors contribution:**

The authors of this scientific article contributed as follows:

Marzieh Asgari-Targhi: Conceptualized and implemented the philosophical framework and theoretical approach, providing the foundational ideas for the article. She wrote and reviewed the manuscript.

Amene Asgari-Targhi: Contributed to the development of the argument specifically in presenting a comparative exposition of mathematical and statistical modeling, mapping mechanistic causality with mathematical modeling. She was responsible for the applied mathematical modeling in the biological case study, illustrating how mechanistic causality can be formally represented and investigated, and how it contrasts with statistical approaches centered on difference-making.

Mahboubeh Asgari-Targhi: Contributed the solar physics case study, experiment, data collection, and analysis. Provided the conceptual and philosophical interpretation of causality applied to the solar corona case study.

Ned Hall: Offered valuable critiques, ensuring overall coherence in the articulation of philosophical themes. He is responsible for the causal interpretation underpinning this paper.

All four authors approved the final version of the manuscript. This multidisciplinary work reflects a genuinely collaborative effort, with each author making significant contributions to the development of ideas, analysis, and writing. While the project benefited from diverse disciplinary perspectives, the core arguments and structure of the manuscript were shaped collectively through ongoing dialogue and joint refinement.